\documentclass[runningheads]{llncs}

\usepackage[T1]{fontenc}
\usepackage{graphicx}
\usepackage{booktabs}
\usepackage{float}
\usepackage{hyperref}
\usepackage{orcidlink}
\usepackage{graphicx}
\begin{document}
\title{Teacher Professional Development on WhatsApp and LLMs: Early Lessons from Cameroon}
%
%

\author{
Vikram Kamath Cannanure\inst{1}\thanks{This paper builds on work accepted at AIED 2026. Please cite the original version when referring to this work.} ,\orcidlink{0000-0002-0944-7074} \and
Bruno Yinkfu\inst{2} \and
Douglas Bryan\inst{2} \and
\\Mati Amin\inst{3} \and
Ingmar Weber\inst{1}\,\orcidlink{0000-0003-4169-2579}
}

\authorrunning{V. Cannanure et al.}
\titlerunning{Teacher Professional Development on WhatsApp and LLMs...}

\institute{
Saarland University, Germany 
\and
B2THEWORLD, United States of America, 
\and
Chatac.ai, United States of America \\
{*\href{mailto:cannanure@cs.uni-saarland.de}{\nolinkurl{cannanure@cs.uni-saarland.de}}}
}

\maketitle              
\begin{abstract}
AI in education is commonly delivered through web-based systems such as online forms and institutional platforms. However, these approaches can exclude teachers in low-resource contexts, where everyday mobile platforms like WhatsApp serve as primary digital infrastructure. To address this gap, we present a field pilot in Cameroon that deploys a WhatsApp-based chatbot with LLM-supported content for teacher professional development (TPD), compared with an online form baseline. The system was evaluated through a mixed-methods study with 47 primary school teachers, integrating quantitative measures with qualitative insights from interviews and participant feedback. Results show that the chatbot was rated higher in perceived usability and overall experience, while learnability remained comparable. These improvements were driven by platform familiarity, low interaction overhead, and the modular structure of LLM-supported content, but were constrained by connectivity limitations, prepaid data costs, and multilingual needs (English/French). Building on these findings, we outline design directions for multilingual, culturally grounded interaction and for supporting prompting and reflection in AI use. We also advocate for teachers’ long-term aspirations through goal-setting and scenario-based learning. More broadly, this work points to \emph{Thoughtful} AI that supports reflection, relevance, and sustained professional growth.

\keywords{Teacher professional development \and AI in education \and WhatsApp \and Equity \and Global South}
\end{abstract}

\section{Description of the AIED implementation practice}

\subsection{Study Context}

\paragraph{Background and Site Context} Teachers in Cameroon operate in a multilingual, mobile-first, and resource-constrained educational environment. This context is shaped by the country’s bilingual education system in English and French and regional cultural diversity \cite{anchimbe2006functional,echu2004language}. These conditions are coupled with uneven digital infrastructure: while mobile phones are widespread, reliable internet remains inconsistent, especially outside urban areas \cite{itu2023facts,gsma2023mobile}. As a result, teachers rely on prepaid data, face frequent network interruptions, and have limited access to institutional systems or personal computers—reflecting broader mobile-first, intermittent connectivity patterns across sub-Saharan Africa \cite{cannanure_we_2022,poon2019engaging}.

Despite these constraints, WhatsApp has become a central communication infrastructure in schools. This platform is widely used for administrative coordination, peer support, and informal professional learning \cite{DIA,varanasi_how_2019,motteram_whatsapp_2020}. In this context, WhatsApp functions not just as a messaging tool but as an embedded socio-technical infrastructure supporting everyday teaching and collaboration \cite{nelimarkka2021facebook,varanasi_tag_2021,cannanure_we_2022}. This role makes it a strong candidate for embedding AI-supported educational tools into teachers’ existing practices.

\begin{figure}[H]
    \centering
    \includegraphics[width=\linewidth]{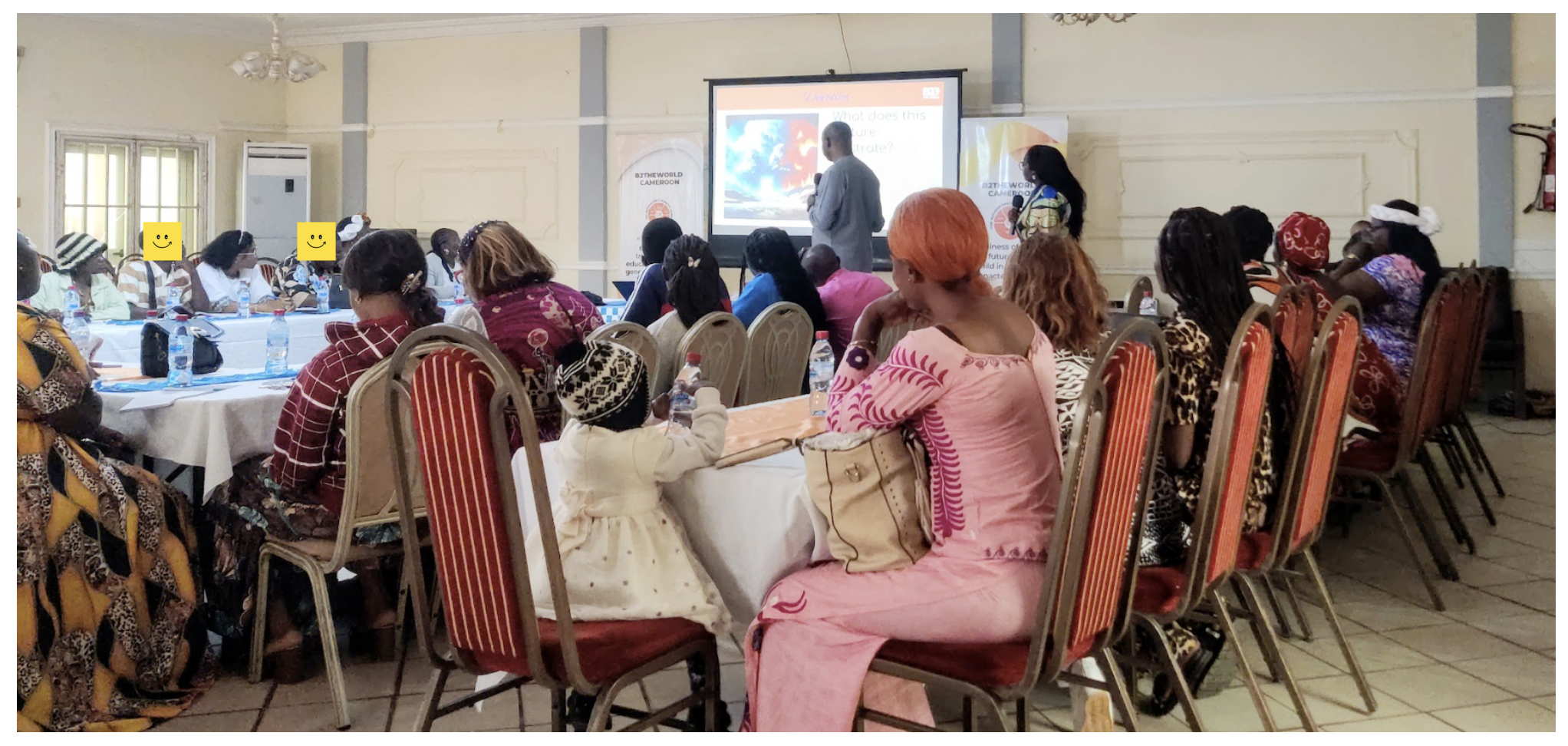}
    \caption{In-person teacher professional development workshop conducted as part of the study. Teachers participated in facilitated group discussions and reflective activities that complemented the WhatsApp-based chatbot learning modules.}
    \label{fig:pd_workshop_cameroon}
\end{figure}

\paragraph{Partner Organization and Training Program} 
The study was conducted in collaboration with an international organization that provides in-service TPD for primary school teachers across multiple regions of Cameroon. The organization works through sustained partnerships with school leaders and teachers, delivering in-person workshops complemented by ongoing follow-up and support activities.

This study was conducted during a teacher training workshop in September 2025 (see Fig.~\ref{fig:pd_workshop_cameroon}). This workshop took place over one week in Bertoua, in Cameroon’s eastern region, with sessions delivered primarily in English and supported by live French translation. A total of 55 primary school teachers from six private schools participated. Within this workshop setting, a WhatsApp chatbot was introduced as an experimental module, with approximately 30 minutes of guided use each day.

\subsection{WhatsApp chatbot}

\begin{figure}[t]
    \centering
    \includegraphics[width=0.85\linewidth]{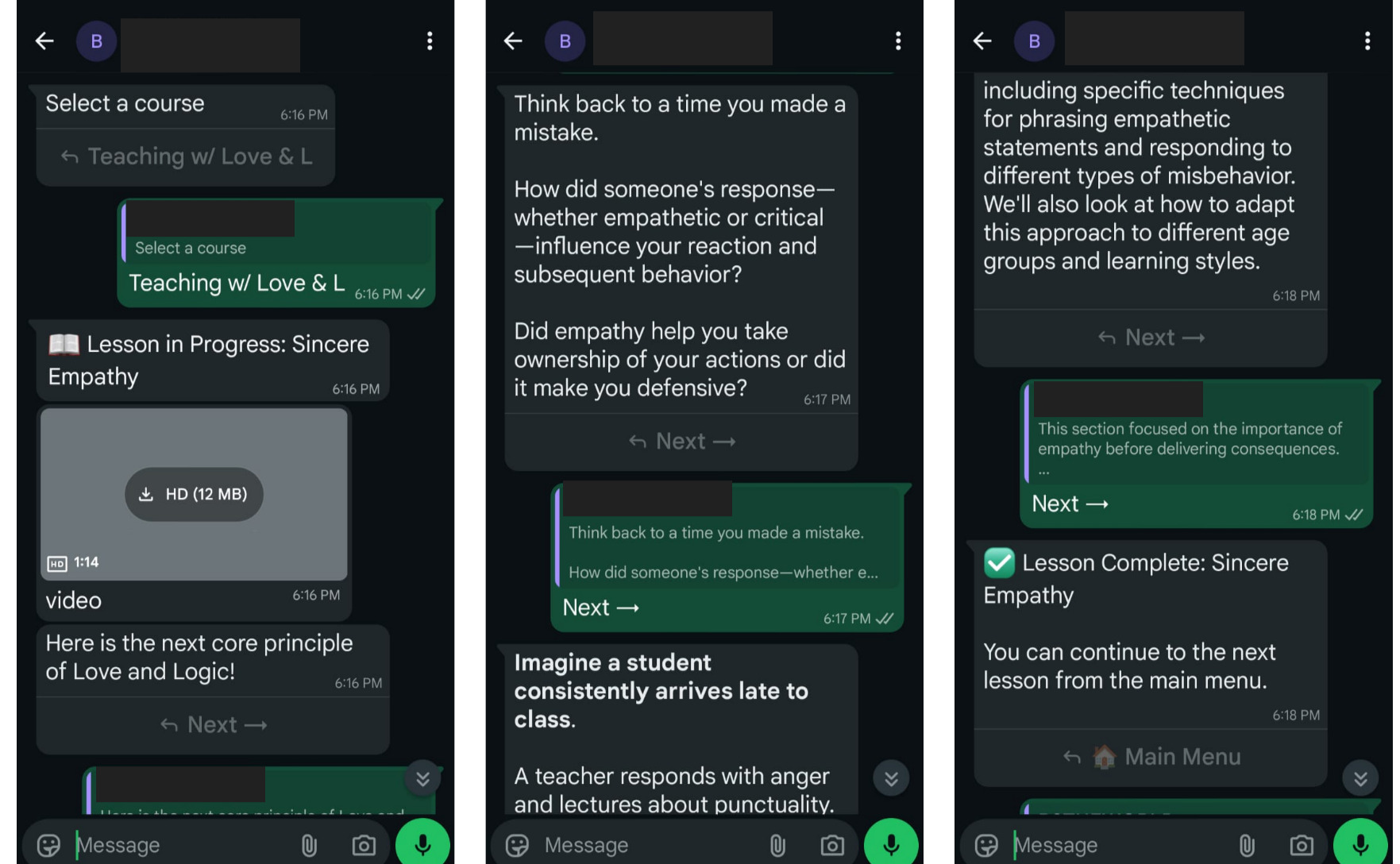}
\caption{Example WhatsApp chatbot lesson flow used during the teacher professional development pilot. Screenshots show (left) course selection and multimedia lesson content; (middle) a reflection prompt embedded in the chat interaction; and (right) lesson completion and navigation to subsequent modules.}
    \label{fig:whatsapp_chatbot_lesson_flow}
\end{figure}

The chatbot delivered TPD through short, structured, conversational modules on WhatsApp (Fig.~\ref{fig:whatsapp_chatbot_lesson_flow}). These modules guided teachers through navigation and pacing (menus and step prompts), presented micro-content (text and occasional multimedia), and prompted reflection by connecting ideas to classroom practice. This content was developed using an LLM-supported dashboard for NGO facilitators (see Appendix Fig.~\ref{fig:dashboard_system}), including drafting explanations, generating examples, and creating variations of reflection prompts, with all outputs reviewed by the NGO facilitator. Some of the later content was generated using the LLM dashboard, while the initial modules were manually created using support from online LLMs. The chatbot interaction and delivery remained menu-based and scripted on WhatsApp. Within WhatsApp, teachers engaged by selecting menu options or entering responses.

The chatbot content was aligned with the partner NGO’s ongoing TPD curriculum. This curriculum drew on established pedagogical approaches rather than bespoke technical content. These approaches included \textit{Teaching with Love \& Logic}, which emphasizes relationship-centered classroom management and shared responsibility \cite{fay1995love}; \textit{Play-Based Learning}, which supports child-centered, activity-driven instruction in early classrooms \cite{hirsh2009play,weisberg2013guided}; and \textit{The Stoplight Approach}, which helps teachers interpret and respond to students’ emotional states using a simple red--yellow--green model grounded in self-regulation research \cite{denham2012sel,stoplight2026impact}. In addition, one module was manually created to introduce the NGO, helping teachers become familiar with its program, work, and mission. 

\subsection{Data Collection}

We used a mixed-methods approach to study teachers’ experiences during a one-week training program. This study was approved by the Ethics Review Board (ERB) at Saarland University (No. ~25-08-6) and included surveys, chatbot interaction data, school-based group discussions, and semi-structured interviews. 


\paragraph{Procedure and data collection}
Teachers first completed a bilingual (English/French) Google Form pre-test with consent and demographics, along with 20 structured questions, serving as a baseline for comparison with the chatbot. Teachers were then onboarded to the WhatsApp-based chatbot through a facilitated training session that used printed and digital instructions, along with guided, hands-on interaction. During the first few days, teachers completed onboarding and initial activities, followed by additional introductory activities over the next two days. After this use period, participants completed a primarily paper-based post-training survey ($n=47$, 39 female, 8 male) to capture perceptions of usability, learnability, challenges, and anticipated future use. To complement the survey, brief school-based group discussions were held across six schools (two representatives per school, English/French), focusing on experiences, benefits, and challenges. Finally, semi-structured interviews conducted a few weeks after the training with a subset of teachers provided deeper insights into trust, language, access, and expectations for future AI-supported professional development. A summary of all data sources and their purposes is shown in Table~\ref{tab:data}.

\begin{table}[H]
\centering
\caption{Summary of Data Collected}
\label{tab:data}
\begin{tabular}{p{3.5cm} p{3cm} p{6cm}}
\toprule
\textbf{Data Source} & \textbf{Participants} & \textbf{Purpose} \\
\midrule
Online form & 40 teachers & Baseline comparison with chatbot; initial experience with web-based interface \\
Chatbot interaction logs & 47 teachers & Patterns of use, completion, and interaction flow \\
Post-training survey & 47 teachers & Metrics for usability, learnability, challenges, and future use \\
School-based group discussions & 6 schools & Collective reflections and improvement suggestions \\
Semi-structured interviews & 14 teachers & In-depth perspectives on AI, trust, and equity \\
\bottomrule
\end{tabular}
\end{table}

\subsection{Data Analysis}
We analyzed the data using a mixed-methods approach to understand teachers’ experiences. This analysis combined quantitative survey data with qualitative feedback from discussions and interviews~\cite{creswell2013steps}. The quantitative data were analyzed using descriptive statistics and paired tests appropriate for this pilot. The qualitative data, including open-ended responses, group discussions, and interviews, were analyzed thematically using inductive coding~\cite{charmaz2008grounded}. These themes focused on patterns related to access, language, infrastructure, and perceptions of AI-supported professional development. The findings were triangulated across methods to strengthen interpretation.

\subsection{Technology access, WhatsApp use, and prior AI exposure.}

\paragraph{Mobile-first access and WhatsApp use}
Teachers primarily relied on mobile phones and WhatsApp for communication and professional learning. This mobile-first environment was reflected in device access: 87.5\% (42/48) owned a personal smartphone, 4.2\% (2/48) shared a phone, and 8.3\% (4/48) reported no phone access. These access patterns were accompanied by widespread WhatsApp use: 77.1\% (37/48) reported daily use and 10.4\% (5/48) reported weekly use. Together, these findings show that WhatsApp was already embedded in teachers’ everyday routines prior to the intervention.

\paragraph{LLM use through embedded and informal channels}
Teachers had prior exposure to LLM-based tools, but this use was informal and uneven. This exposure was shaped by platform integration, with a majority 62.5\% (30/48) identifying Meta AI—embedded within Facebook or WhatsApp—as their primary tool. Teachers attributed this preference to immediacy and familiarity, noting that \emph{“it’s already there...you just enter and type your question”} (T5). In contrast, 16.7\% (8/48) preferred ChatGPT, which they described as easy to install and use independently: \emph{“you just download the application, type your question, and it gives you the answer”} (T5). Only 2.1\% (1/48) reported no prior AI use. Language preferences further shaped access: 60.4\% (29/48) were more comfortable in English, and 39.6\% (19/48) in French. These patterns suggest that teachers prefer AI tools that are lightweight, mobile, and embedded within existing communication platforms, shaping how AI-supported TPD can be accessed and adopted.

\subsection{Prior work: AI-Supported Teacher Professional Development and Reflective Practice}

Scaling effective teacher professional development (TPD) remains a challenge in low-resource contexts. TPD is critical for instructional quality and teacher confidence, yet many programs rely on one-off workshops or lightweight web-based tools (e.g., online forms) that offer limited support for sustained reflection and feedback \cite{mulkeen2005teachers,bennell2004teacher}. Mentoring and peer-learning approaches, including virtual communities of practice, have shown initial impact \cite{Study4}, but they are difficult to scale where time, infrastructure, and expert facilitation are limited \cite{Study1}. Recent work suggests AI could help address these constraints by providing on-demand reflection support, feedback, and access to instructional resources \cite{tan2024ai,li2024ai}. However, most AI-in-education systems are student-focused and evaluated in well-resourced settings with stable connectivity. As a result, little is known about AI-supported TPD in mobile-first, low-connectivity, bilingual contexts such as Cameroon.

In many African contexts, WhatsApp already serves as a practical infrastructure for teacher coordination, peer support, and informal professional learning \cite{mary_mendenhall_expanding_2018,motteram2019resilience,DIA}. This widespread use makes WhatsApp a natural platform for extending TPD, but existing practices are often unstructured and require ongoing facilitation \cite{mary_mendenhall_expanding_2018,nelimarkka2021facebook}. Embedding structured, AI-supported instruction with WhatsApp may provide more consistent scaffolding at scale \cite{Grudin2019,DIA,cannanure_we_2022}. However, there is limited field evidence on how AI-supported WhatsApp chatbot approaches compare with conventional web-based tools such as online forms, and on the benefits and challenges they entail in low-resource settings. 

\paragraph{Research Questions}
Drawing on survey data, system usage, group discussions, and interviews, we address the following research questions:

\begin{itemize}
    \item \textbf{RQ1:} How do teachers’ experiences differ when completing training activities using a WhatsApp-based chatbot versus a conventional online form?

    \item \textbf{RQ2:} What are the benefits and challenges of using a WhatsApp-based chatbot for TPD in low-resource school contexts?
\end{itemize}

\section{Reflection on challenges and opportunities}

We report findings from surveys, system logs, and semi-structured interviews, organized around our research questions.

\subsection{RQ1: Comparing Chatbot and Online Form Experiences}

\begin{table}[H]
\centering
\caption{Comparison of Means: Chatbot vs. Online Form Experiences (RQ1)}
\label{tab:rq2_comparison}
\begin{tabular}{lcccc}
\toprule
\textbf{Measure} & \textbf{Chatbot} & \textbf{Form} & \textbf{\emph{t}(43)} & \textbf{\emph{p}-value} \\
\midrule
Usability & 7.82 & 7.33 & 2.05 & \textbf{.046*} \\
Learnability & 7.00 & 6.76 & 1.27 & .212 \\
Combined Scale & 7.41 & 7.03 & 2.08 & \textbf{.044*} \\
\bottomrule
\end{tabular}

\vspace{0.5em}
\footnotesize{Note: *$p < .05$. Effect sizes (Cohen’s \emph{d}) were small to moderate for usability (\emph{d} = 0.37) and the combined scale (\emph{d} = 0.32), and very small for learnability (\emph{d} = 0.16).}
\end{table}

We compared teachers’ experiences between the chatbot and the online form across usability, learnability, and overall experience. These measures demonstrated acceptable internal consistency. The usability scale showed good reliability (Cronbach’s $\alpha = .81$), the learnability scale showed acceptable reliability ($\alpha = .74$), and the combined scale showed good reliability ($\alpha = .83$).

\paragraph{Usability.}
The chatbot was rated as significantly more usable than the online form. This difference is reflected in higher mean usability scores for the chatbot ($M = 7.82$) compared to the online form ($M = 7.33$), and was statistically significant, $t(43) = 2.05$, $p = .046$, with a small–to–moderate effect size (Cohen’s $d = 0.37$). These differences were supported by qualitative data, which suggests that familiarity with WhatsApp and the chatbot’s low interaction overhead contributed to higher usability. As one teacher noted, \emph{``I use WhatsApp every day, so it was easy''} (T5). In contrast, the online form was often described as requiring additional explanation or assistance, particularly at the start.

\paragraph{Learnability.}
Learnability did not differ significantly between the chatbot and the online form. This is reflected in similar mean scores for the chatbot ($M = 7.00$) and the online form ($M = 6.76$), $t(43) = 1.27$, $p = .212$, with a very small effect size ($d = 0.16$). These findings were supported by qualitative accounts indicating that both tools could be learned successfully, but that they relied on different forms of support: step-by-step prompts within the chatbot versus facilitator or peer support for the online form.

\paragraph{Combined experience.}
The chatbot showed a significantly better overall experience than the online form. This is reflected in higher combined scores for the chatbot ($M = 7.41$) compared to the online form ($M = 7.03$), $t(43) = 2.08$, $p = .044$, with a small–to–moderate effect size ($d = 0.32$). These differences were supported by qualitative responses, where teachers emphasized the chatbot’s speed, clarity, and familiarity, and expressed greater confidence in using it independently. In contrast, the online form was more often associated with challenges related to connectivity, language, and the need for assistance.

\subsection{RQ2: Benefits and Challenges of WhatsApp-Based Chatbot for Professional Development}

\paragraph{Familiarity and low access barriers.}
The chatbot was easy to adopt because it was built on tools teachers already used. Teachers described joining the system by saving a number and following step-by-step instructions, without needing to learn a new platform. As one teacher explained, \emph{``everything was clear and directed step by step… before you leave the first stage, they will tell you what to do''} (P2). Others emphasized language and routine familiarity, noting, \emph{``it was easy because the language was one I could understand, so I could follow what was asked and respond''} (Survey), and \emph{``I use WhatsApp every day, so it was easy''} (T5). These accounts suggest that aligning with existing practices reduced onboarding effort and made participation feel natural rather than technical.

\paragraph{Clear structure and support for teaching practice.}
Teachers used the chatbot to support lesson preparation and classroom practice. Building on familiar WhatsApp interactions, the chatbot provided structured guidance through short explanations, reflection prompts, and concrete examples that could be directly applied in classrooms. Teachers also valued access to examples, videos, and step-by-step guidance at their own pace. As one participant noted, \emph{it facilitates learning… it makes the lessons easier''} (T5), while another shared, \emph{if I am blocked somewhere, it can help me’’} (P2). Teachers also mentioned early opportunities of practice-linked reasoning, such as connecting module concepts to classroom activities. Several teachers described returning to the chatbot as a reference for teaching strategies; for example, one noted it \emph{``serves as a reference where I can return to teaching strategies when I need them’’} (T12).

\paragraph{Connectivity and infrastructure constraints.}
Connectivity and data limitations frequently disrupted the chatbot's use. Teachers described slow loading times, interrupted sessions, and difficulty completing activities when network coverage was weak. As one participant summarized, \emph{``the only difficulty is the network''} (T12), while another noted, \emph{``those who never had a connection were struggling to connect''} (T5). These accounts suggest that participation was shaped less by system usability and more by external infrastructure conditions such as data availability and signal strength.

\paragraph{Language and interaction limitations.}
Language barriers and rigid interaction flows limited the chatbot’s accessibility and flexibility. Teachers noted that English-only content excluded some Francophone colleagues; as one participant observed, \emph{``we had those other French (speaking) brothers and sisters that were not understanding the English very well''} (T14), highlighting the need for French support. Participants also described the chatbot’s fixed navigation as restrictive. As T4 explained, \emph{``it always brings me back there… you have to keep at it until it happens.''} Teachers expressed a need for more flexible and interactive features, including personalized feedback and the ability to ask open-ended questions.

\section{Future Steps and Implementation Needs}

Building on pilot findings from a one-week deployment ($n=47$) across comparative experience (RQ1) and reported barriers (RQ2), we outline future design directions for AI in rural communities informed by teachers’ voices.


\paragraph{From Prompting to Reflection: Supporting Thoughtful AI Use.}
Teachers emphasized that effective use of LLMs requires careful prompting and active effort. Many described iteratively refining their questions when responses were unclear or not locally relevant. As one teacher noted, \emph{``it depends on how the question is asked''} (T5), explaining the importance of prompt formulation. At the same time, teachers expressed concern that new AI users may be misled by responses that appear correct but are not appropriate in context. They also highlighted the risk of reduced reflection when AI is used uncritically. As one teacher explained, \emph{``it (AI) can make people think less, because answers come so quickly… it should encourage teachers and learners to reflect, not only to copy ready-made answers''} (T8). These reflections point to the need for scaffolding both prompting practices and critical engagement with AI. Future work can support teachers in developing prompting strategies and reflective use of LLMs in classroom contexts, building on emerging evidence from AI-supported learning interventions (e.g., large-scale deployments such as Microsoft Copilot–based tutoring in Nigeria \cite{worldbank2024chalkboards}). Thus, designing for \emph{Thoughtful} AI that encourages reflection rather than passive consumption.

\paragraph{Expanding language support and cultural sensitivity.}
Language and cultural misalignment limited both access and relevance in this bilingual context. Teachers noted that English-only interaction excluded some users and emphasized the need for support in French and regional languages commonly used in their communities.  Teachers also highlighted gaps in how LLMs reflect local cultural contexts. As one teacher explained, \emph{``I do not feel that AI tools really consider our culture in Cameroon… when I researched ‘inclusive education’, I expected more focus on children with disabilities, but the tool focused on issues like transgender inclusion, which felt like it was imposing another culture''} (T9). In response, teachers expressed interest in contributing to community-driven datasets that incorporate locally relevant examples and perspectives. These findings point to the importance of participatory data collection and localization—such as initiatives like Digital Umuganda in Rwanda for building voice datasets \cite{mozilla2023kinyarwanda}—to support the design of culturally grounded AI systems for TPD.

\paragraph{Supporting long-term aspirations and professional growth.}
Teachers saw AI as a tool not only for immediate tasks but also for supporting their long-term professional goals \cite{cannanure_we_2022,Study1}. Teachers described how AI could help them plan future pathways, develop new skills, and explore opportunities such as pursuing further education or becoming trainers. As one teacher shared, \emph{``AI can help me by giving clarity and guidance about my vision and objectives… what opportunities a master’s degree might open and what steps I should take''} (T7). Beyond pedagogy, teachers highlighted how AI could connect them to broader knowledge beyond their local context. For example, one teacher described how access to \emph{``knowledge from many cultures and contexts''} (T12) could support their goal of becoming a trainer and help them think through potential research directions. 

Future work can build on this by designing systems that support long-term TPD with AI. For example, AI-powered tools could include adaptive goal-setting features \cite{Can_DBR} alongside virtual roles or scenario-based interactions \cite{Study4} (e.g., simulating being a trainer), helping teachers for future opportunities.

\paragraph{Scaling deployment and strengthening evaluation.}
Sustained impact will require moving beyond short-term pilots to long-term, real-world deployment. Building on this initial study, future work can focus on longitudinal implementations \cite{Study4} to assess sustained engagement and changes in teaching practice over time. While LLMs in this work primarily supported content creation and refinement, the next step is to more tightly integrate AI into teacher-facing interactions.  This shift points toward the design of \emph{Thoughtful} AI systems that support reflection, contextual adaptation, and sustained engagement in everyday teaching practice.

%
%
%
%

\bibliographystyle{splncs04}
\bibliography{refs_clean}

\section{Appendix}

\begin{table}[H]
\centering
\caption{Participant demographics and technology use ($n=14$). 
Teaching language and WhatsApp use were self-reported in the post-training survey. 
Chatbot messages represent the total number of messages sent during the study period.}
\label{tab:participants}
\resizebox{\linewidth}{!}{
\begin{tabular}{c c c c c c c}
\toprule
\textbf{PID} &
\textbf{Gender} &
\textbf{Age} &
\textbf{Teaching Language} &
\textbf{Teaching Level} &
\textbf{WhatsApp Use} &
\textbf{Chatbot Messages} \\
\midrule
T1  & F & 24 & English  & Primary  & Daily  & 238 \\
T2  & F & 45 & French   & Primary  & Daily  & 173 \\
T3  & F & 32 & French   & Primary  & Daily  & 85  \\
T4  & F & 35 & French   & Nursery  & Weekly & 73  \\
T5  & F & 30 & English  & Nursery  & Daily  & 63  \\
T6  & F & 27 & English  & Nursery  & Daily  & 56  \\
T7  & F & 21 & English  & Primary  & Daily  & 55  \\
T8  & F & 30 & English  & Primary  & Daily  & 31  \\
T9  & F & 36 & English  & Primary  & Weekly & 29  \\
T10 & M & 32 & French   & Primary  & Daily  & 26  \\
T11 & M & 24 & English  & Primary  & Daily  & 23  \\
T12 & M & 26 & English  & Primary  & Weekly & 15  \\
T13 & F & 34 & English  & Primary  & Daily  & 26  \\
T14 & F & 29 & English  & Primary  & Daily  & 24  \\
\bottomrule
\end{tabular}
}
\end{table}

\begin{figure}[t]
    \centering
    \begin{minipage}[t]{0.49\linewidth}
        \centering
        \includegraphics[width=\linewidth]{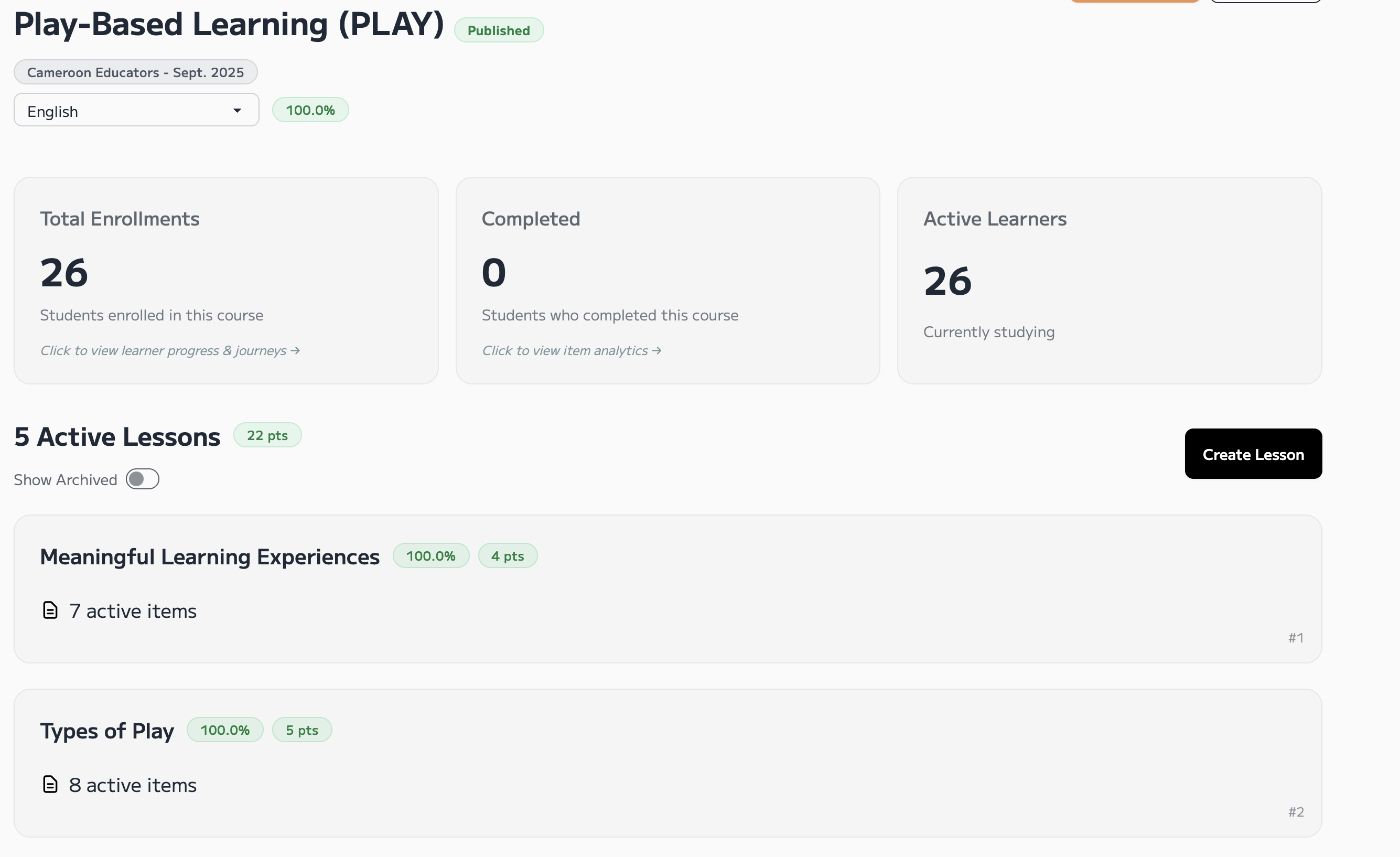}
        \\[2pt]
        {\small (a) Course dashboard showing enrollments, active learners, and lesson modules.}
    \end{minipage}
    \hfill
    \begin{minipage}[t]{0.49\linewidth}
        \centering
        \includegraphics[width=\linewidth]{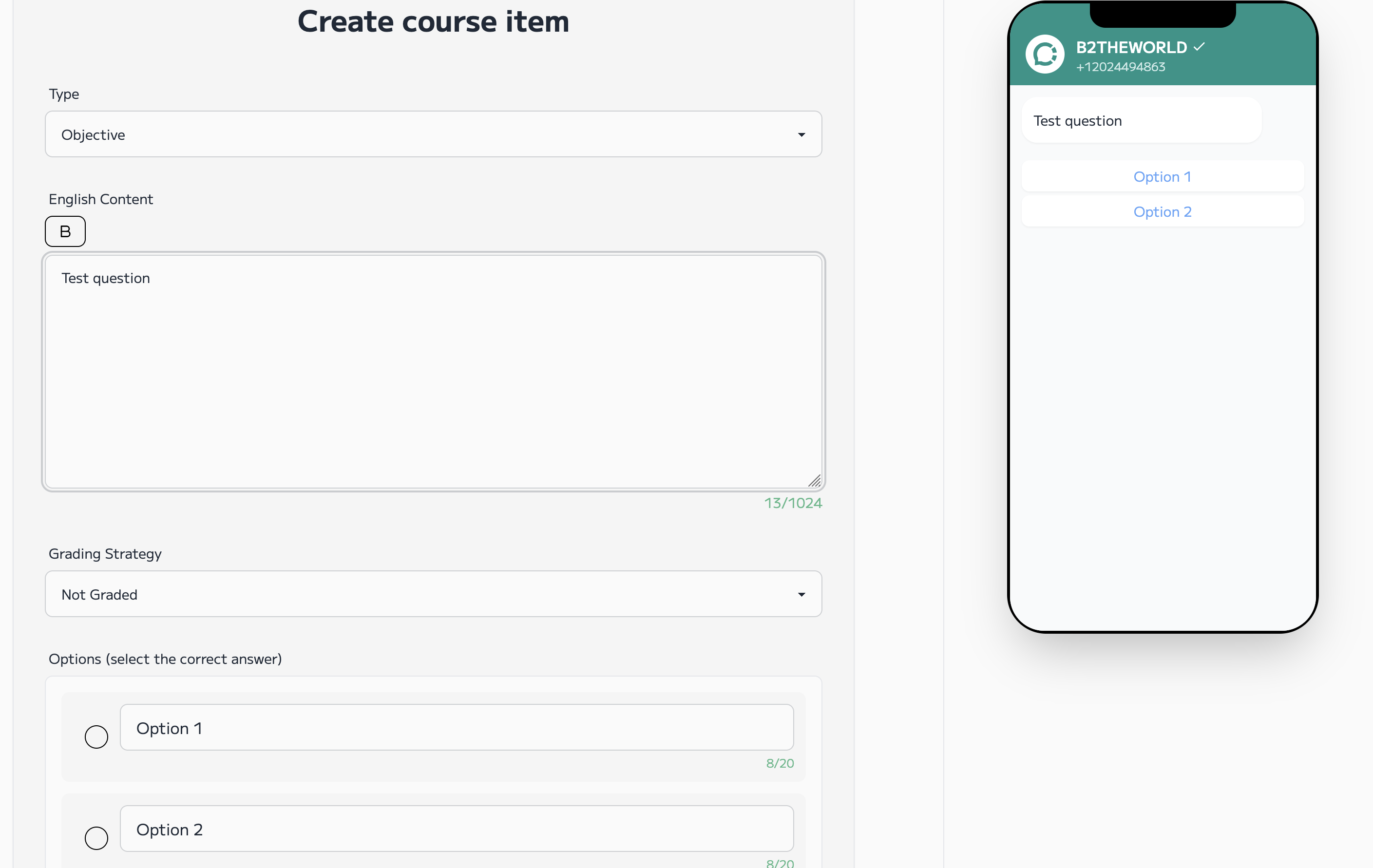}
        \\[2pt]
        {\small (b) Authoring interface for creating chatbot learning items and previewing WhatsApp delivery.}
    \end{minipage}
    \caption{Instructor-facing dashboard for managing WhatsApp-based professional development content. Left: course analytics and lesson structure. Right: item creation interface with real-time WhatsApp preview.}
    \label{fig:dashboard_system}
\end{figure}

\section*{Acknowledgments}
VKC and IW are supported by the Alexander von Humboldt Foundation. The authors thank their partners in Cameroon for facilitating the teacher training program and supporting data collection, and the participating teachers for their time and insights.

\subsection*{Disclosure of Interests}
Vikram serves as an advisor to Chatac.ai, a startup working on conversational AI systems. Mati Amin is affiliated with Chatac.ai. B2THEWORLD has a client relationship with Chatac.ai, which collaborated on the deployment described in this study. These relationships did not influence the study design, data collection, analysis, or reporting of results. The remaining authors declare that they have no competing interests.

\end{document}